\documentclass[twocolumn,showpacs,preprintnumbers,amsmath,amssymb]{revtex4}
\usepackage{graphicx}
\usepackage{dcolumn}
\usepackage{bm}

\begin{document}

\title{Electronic structure Studies of CeAgSb$_{2}$}

\author{T. Jeong}

\affiliation{
Department of Physics, University of California, Davis, California 95616
}

\begin{abstract}
The electronic band structure of CeAgSb$_{2}$ has been calculated using 
the self-consistent full potential nonorthogonal local orbital minimum
basis scheme based on the density functional theory.  We investigated 
the electronic structure with the spin-orbit interaction and on-site 
Coulomb potential for the Ce-derived 4f orbitals to obtain the correctly 
ground state of CeAgSb$_{2}$.

\end{abstract}

\pacs{71.10.Hf, 71.18.+y, 72.20.Eh, 75.30.Mb}

\maketitle

\section{Introduction}
The tenary compounds CeXSb$_{2}$ (where X=Au, Ag, Cu, Ni, or Pd) have 
attracted considerable attention recently because they exhibit 
a large variety of physical properties such as heavy fermion, Kondo 
insulating, anisotropic transport and magnetic ordering behavior.
The various physical properties depend on the hybridization between 
f electrons and conduction electrons, which is characterized by a 
Kondo temperature $T_{K}$. With a small hybridization the system exhibits
a local magnetic moment and orders magnetically, which can be 
described by the Ruderman-Kittel-Kasuya-Yosida (RKKY) interaction. 
By increasing the hybridization, the Kondo effect increases 
and the ordered magnetic moment decreases gradually. With further 
increase of the hybridization, the system will go into a heavy 
fermion or intermediate valence regime.

CeAgSb$_{2}$ is a Ce-based Kondo lattice system with 
low-temperature ferromagentic ordering.
Sologub {\it et al.} \cite{sologub} reported this compound 
has weak ferromagnetic oder
in polycrystalline samples with a net ferromagnetic moment of 
0.15 $\mu_{B}$/Ce at 5 K.
Several different groups investigated the magnetic properties of 
CeAgSb$_{2}$ with conflicing results for the magnetic ground state.
\cite{takeuchi, houshiar, thornton}.
In particular, for single crystal smaples, the magentization of
CeAgSb$_{2}$ is anisotropic\cite{myers}.
In this case the magnetization for the magnetic field applied
parallel to the c-axis shows a ferromagnetic order with a moment of
0.37 $\mu_{B}$/Ce above 0.025 T at 2 K.
For H$\perp$c, the magnetization increases nearly linearly 
to 1.2 $\mu_{B}$/Ce below 30 kOe and then remains nearly constant 
for higher fields, possibly indicating the presence of a metamagnetic 
transition. Above 200K, the inverse susceptibility is linear to the 
Curie-Weiss law giving an effective moment of 2.26  $\mu_{B}$/Ce. 
In zero field, the resistivity of CeAgSb$_{2}$ increases rapidly from 
1.16 $\mu \Omega $cm at 2K to a maximum of 88.1 $\mu \Omega $cm 
at 18.2 K. At 9.7K, a sharp change in the slope of the zero field 
resistivity is observed, consistent with a loss of spin-disorder 
scattering associated with the magnetic ordering as well as 
the possible suppression of the Ce hybridization due to the ferromagnetic 
ordering. 
At higher temperatures, the resistivity determines a broad 
local minimum near 150K. 
The temperature-dependent resistivity of CeAgSb$_{2}$  is typical 
of a Kondo lattice system. Preliminary inelastic neutron scattering 
measurements indicate a Kondo temperatur T$_{K}$ between 60 and 80K
\cite{thornton}. 
The magnetoresistence of CeAgSb$_{2}$ is either positive or negative,
depending on the temperature and orientation of the applied field. 
This complexity is due to competing contributions from the Ce hydridization, 
magnetic ordering, and the electronic structure of the compound.
The magnetoresistence of a paramagnet is expected to be negative, since 
the local moments try to align parallel to the applied field, which essentially 
reduces the spin disorder scattering as the field is increased.

In order to understand the electronic and magentic properties of CeAgSb$_{2}$,
we need the electronic band structure studies based on the density 
functional theory.
In this work, the precise self-consistent full potential 
local orbital minimum basis band structure scheme (FPLO) are 
employed to investigate the electronic and magnetic 
properties of CeAgSb$_{2}$ with LDA, LDA+U and fully relativistic shemes.
We consider the effect of magnetism on the band structure and 
compare with experiment. 

\section{Crystal Structure}
CeAgSb$_{2}$ crystallizes in the primitive tetragonal ZrCuSi$_{2}$
structure, which consists of Sb-CeSb-Ag-CeSb-Sb                    
layers along $[001]$ direction. 
This crystal structure is described in detail by Myers {\it et al.}
\cite{myers}.
It belongs to the P4/nmm space group with 
Ce occupying the 2c site, Ag the 2a, and Sb the 2b, 2c sites.
The sites are given, in units of (a, a, c): Ce(0.25, 0.25, 0.26),
Ag(0.75, 0.25,0.5), Sb(0.25,0.25,0.84), Sb(0.75, 0.25, 0).
We used experimental lattice constants, a= 4.36 $\AA$ and c=10.41  $\AA$,
in the calculation described below.
There are two formula units in the primitive cell.


\section{Method of Calculations}

We have applied the full-potential 
nonorthogonal local-orbital minimum-basis (FPLO) scheme within the local 
density approximation (LDA).\cite{koepernik}
In these scalar relativistic calculations we 
used the exchange and correlation potential of Perdew and Wang.\cite{perdew}
Ce $4s,4p,3d$, Ag $5s,5p,4d$ and Sb $4s, 4p$ states were included as 
valence states. All lower states were treated as core states.
We included the relatively extended semicore 4s, 4p, 4d, 4f, 5s, 5p states 
of Ce and
4s, 4p states of Sn and Ag as band states
because of the considerable overlap of these
states on nearest neighbors.
This overlap would be otherwise neglected in our FPLO scheme.
Ce 6p states were added to increase the quality of the basis set.
The spatial extension
of the basis orbitals, controlled by a confining potential $(r/r_{0})^4$, was 
optimized to minimize the total energy. The self-consistent potentials were 
carried out on a k mesh of 18 k points in each direction of the Brillouin zone,
which corresponds to 550 k points in the irreducible zone.
A careful sampling of the Brillouin zone is necessitated by the 
fine structures in the density of states near Fermi level E$_{F}$.

\begin{figure}
\vskip -5mm
\includegraphics[height=8.5cm,width=8.5cm,angle=-90]{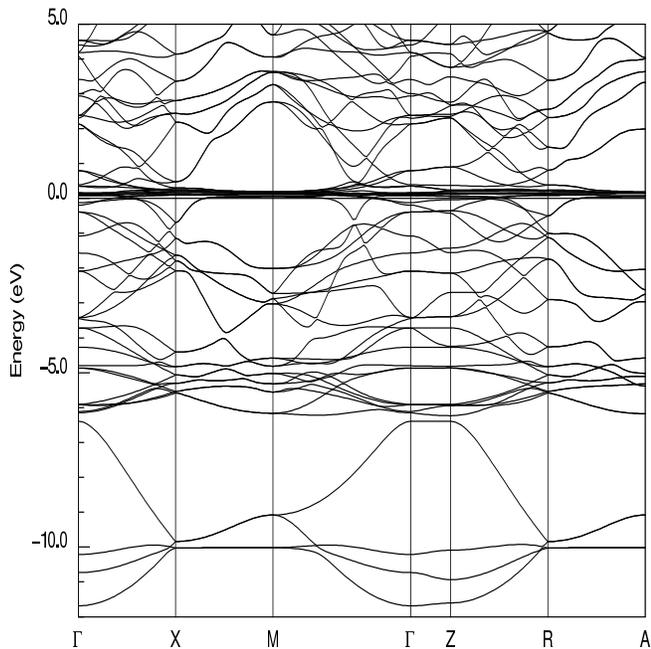}
\caption{The full LDA 
nonmagnetic band structure of CeAgSb$_{2}$ along the symmetry lines.} 
\label{fullband}
\end{figure}


\begin{figure}
\vskip -10mm
\includegraphics[height=8.5cm,width=7.5cm,angle=-90]{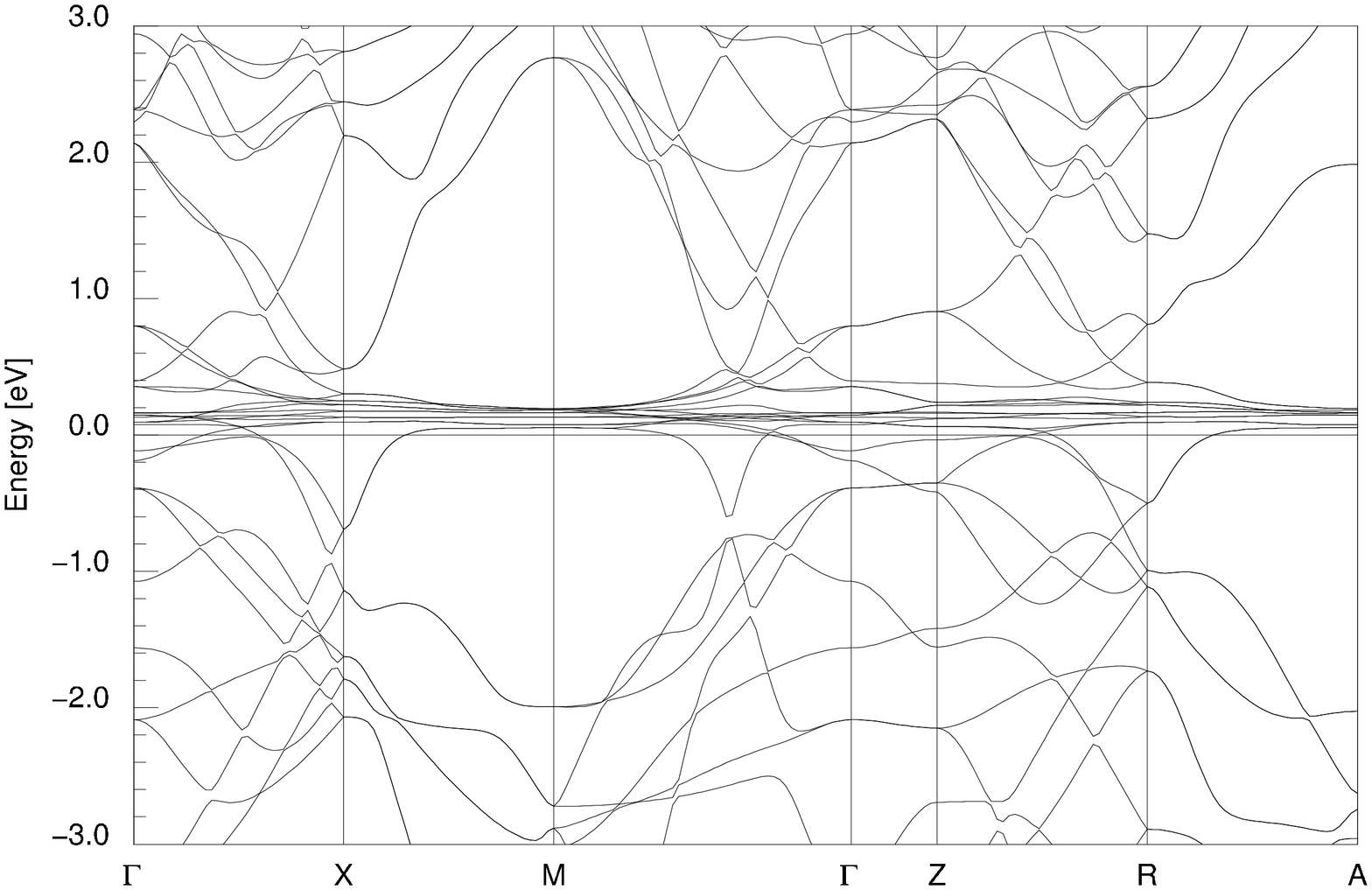}
\vskip -10mm
\includegraphics[height=8.5cm,width=7.5cm,angle=-90]{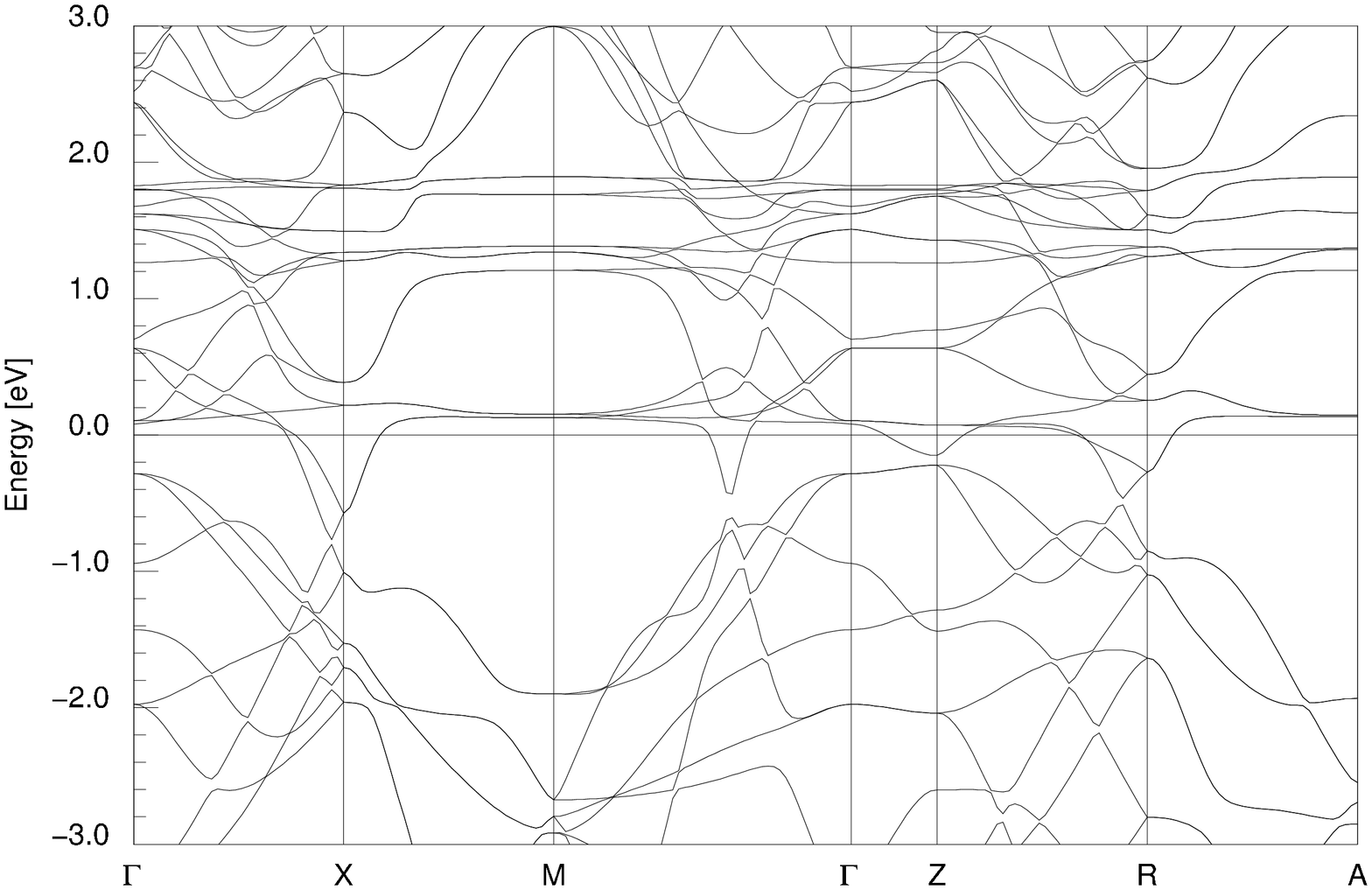}
\vskip -10mm
\includegraphics[height=8.5cm,width=7.5cm,angle=-90]{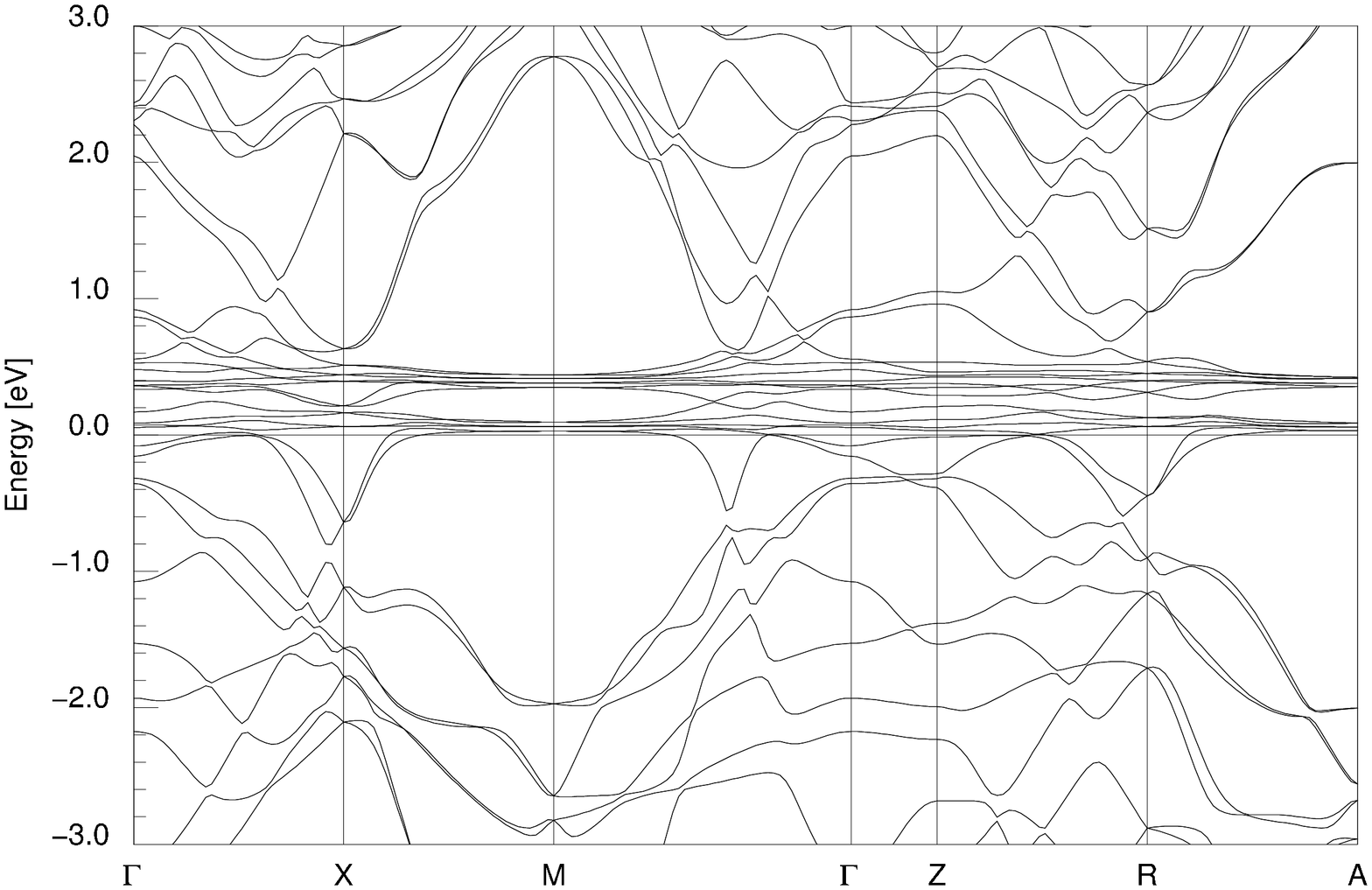}
\caption{
Top panel:The LDA band structure of CeAgSb$_{2}$ along symmetry lines.
The very flat bands near the Fermi level are the Ce 4f bands.
Middle panel:The band structure within the LDA+U scheme showing that
the 4f bands are split in a 1-3-3 fashion from bottom up.
Bottom panel:The fully relativistic band structure of CeAgSb$_{2}$ showing
that spin-orbit coupling splits the 4f states into two manifolds.} 
\label{ldau}
\end{figure}

\section{Results and discussion}

\begin{figure}
\vskip 15mm
\includegraphics[height=8.5cm,width=8.5cm,angle=-90]{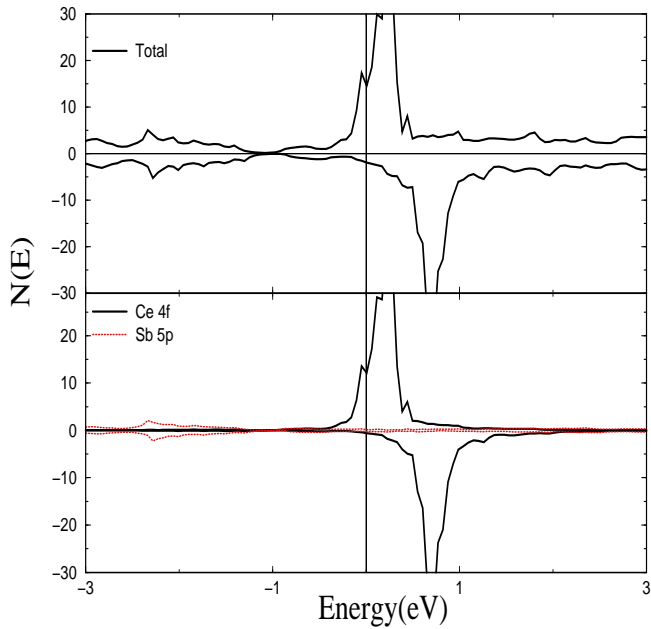}
\caption{
Projected density of states of CeAgSb$_{2}$.
Top panel: total, plotted positively for majority, negatively for minority.
Bottom panel: projection of the Ce 4f and Sb 5p, showing that Ce 4f 
character dominates the states near the Fermi level.} 
\label{dos}
\end{figure}


We first show the full band structures of CeAgSb$_{2}$
within LDA scheme in Fig. \ref{fullband}. 
The Sb 5s bands lie between -11.5eV and -6.5eV.
Between -6.5 and the Fermi level there are of mixed Sb 5p and 
Ag 4d states. 
Those very flat bands near the Fermi level are mainly the Ce-centered 4f 
characters.
A prominent feature of the band structure near $E_{F}$,
besides the 4f bands, is the Ce 5d character which 
hybridizes with the Ce 4f bands.
We also study the on-site atomic-like correlation effects beyond 
LDA by using LDA+U approach in a rotationally invariant, full potential 
implementation\cite{pickett}.
Minimizing the LDA+U total energy functional with spin-orbit coupling(SOC)
treated self-consistently \cite{shick} generates not only the 
ground state energy and spin densities, but also effective one-electron 
states and energies that provides the orbital contribution to the 
moment and Fermi surfaces. The basic difference of LDA+U calculations 
from the LDA is its explicit dependence on the on-site spin and orbitally 
resolved occupation matrices.
The Coulomb potential U and the exchange coupling J for the Ce 4f orbitals
have been chosen to be 7.5 and 0.68 eV respectively. 
The resulting band structure calculated within the LDA+U scheme is shown in 
middle panel of Fig. \ref{ldau}.
We observe that the crystal field splittings
of Ce 4f bands within LDA are quite small
and in fact difficult to identify due to hybridization with itinerant 
bands. From LDA+U, the 4f bands are still very flat but are split
(in a 1-3-3 fashion from bottom up) by some combination of the 
crystal field and the anisotropy of the U interaction by a total 
1.6 eV. 
We also calculated fully relativistic band structure  
to see the spin-orbit coupling effects, which is shown 
in the lower panle of Fig. \ref{ldau}.
As expected spin-orbit coupling splits the 4f states 
into two manifolds, located 0.4 eV and 0.05 eV above the Fermi level,
the 4f$_{7/2}$ and the 4f$_{5/2}$ multiplet respectively.

The atom and symmetry projected densities of states (PDOS) shown 
in Fig.\ref{dos} clarify the characters of the bands.
Because only the DOS distribution near the Fermi level determines 
the magnetic properties, we concentrate our attention upon 
the DOS in the vicinity of the Fermi level. 
At this range, the valence 
states for Ce or Sb atoms are dominated by 4f and 5p electrons, 
respectively, and the contributions from other electrons are negligibly 
small. 
Because of hybridization, they distribute in a wide energy range and extend 
to the unoccupied states, above the Fermi level. These states 
remain almost unchanged when the spin-orbital interaction or on-site 
correlation potential are taken into consideration. So only 
the DOS obtained by LDA calculation for Ce-4f and Sb-5p have been 
plotted. 
From the Fig. \ref{dos}, it can be found that within the LDA calculation, the 
DOS at the Fermi level are larger and they are mainly of Ce-derived 
4f states. The contribution from Ag-derived 4d or Sb-derived 5p states 
is smaller. When the spin-orbital couplilng along the axis (0 0 1) 
is taken into account. 
the 4f orbitals are slightly modified. The spin 
4f states become wider and the energy shift between centers becomes 
larger. This is due to the partial splitting between the degenerate 
4f states. 

Band calculatioin with the LDA framework cannot yield correctly 
magnetic moment for many Ce compounds because of the strong 
correlation interaction between f orbitals. The spin-orbit 
interaction in these systems is sometimes large and the orbital 
contribution to magnetic moment cannot be neglected. 
The calculated magnetic moment of CeAgSb$_{2}$ within LDA scheme 
is 0.61 $\mu_{B}$/Ce. it is mainly form the Ce-derived 4f orbitals, while 
almost no contribution from Ag.
This is in agreement with the fact that the transition metal T, except Mn, 
does not carry magnetic moment in CeTSb$_{2}$ compounds. 
When the on-site correlation potential is added to the Ce 4f 
electron, the degeneracy between the different f orbits would been 
lifted and the Hund's rules dominate the locally occupied 
4f electrons, which yields  
the total magnetic 
moment  0.54 $\mu_{B}$/Ce. 
With fully relativistic scheme we calculated the magnetic moment of the value
0.55 $\mu_{B}$/Ce.
As mentioned above, several groups reported the 
experimental studies on the magnetic moments 
which do not agree with another, 
so further experimental investigations are required.

Density functional calculations are very reliable in calculating 
the instability to ferromagnetism.
The presence of an electronic instability is signaled by a divergence of the
corresponding susceptibility. In the following we study the uniform
magnetic susceptibility using the method of Janak \cite{jan77}.
The uniform magnetic susceptibility of a metal can be written as

\begin{equation}
\chi=\frac{\chi_{0}}{1-N(E_F)I},
\end{equation}

where the numerator stands for the Pauli susceptibility of a gas of non-interacting
electrons proportional to the density of states at the Fermi level $N(E_F)$,
and the denominator represents the enhancement due to electron-electron
interaction. Within the Kohn-Sham formalism of density functional
theory the Stoner parameter $I$ is related to the second derivative of the
exchange-correlation fuctional with respect to the magnetization density.
We have evaluated, within the density functional theory formalism, the Stoner
enhancement of the susceptibility
$\chi=\frac{\chi_{0}}{1-IN(E_{F})}\equiv S\chi_{0} $,
where $\chi_{0}=2\mu_{B}^{2}N(E_{F})$
is the non-interacting susceptibility
and $S$ gives the electron-electron enhancement in terms of the Stoner
constant $I$. We have calculated $I$ using both the Janak-Vosko-Perdew theory \cite{jan77}
and fixed spin moment calculations\cite{mohn}.
The calculated density of states and Stoner parameter normalized per
are $N(E_F)$=5.8 states/eV and $I$=0.19 eV.
This gives $IN(E_{F})=1.1 $, larger than unity, corresponding
to a ferromagnetic instability.




Heavy fermion compound is characterized by a larger electronic 
specific heat coefficient $\gamma$. CeAgSb$_{2}$ is a moderate heavy fermion 
compound with $\gamma= 65$ mJ/K$^{2}$mol\cite{inada}.
The large specific heat coefficient of CeAgSb$_{2}$ compound could not 
be yielded by our band calculation. This can be seen from the 
calculated electronic structure. It can be found that the toal 
number of DOS at the Fermi level is about 5.8 states/eV, which corresponds 
$\gamma_{b}= 13.6$ mJ/K$^{2}$mol and underestimate the experiment value by 
a factor of 4.8. 
The discrepancy between the band calculation and experiment for specific 
heat coefficient is attributed to the 
formation of quasiparticle. There is exchange interaction 
J between the local f and the conduction electrons in CeAgSb$_{2}$.
The ground state of Ce compound is determined by the competition 
of the Kondo and indirect RKKY interaction. With a large J, the Kondo 
coupling becomes strong and the system located at the borderline 
of magnetic-nonmagnetic transition. 
The exchanging interaction between the 
local f electron and the conduction electrons will result in the formation 
of quasiparticle. It has a larger mass compared with bare 
electron and the enhancement of mass increases with the increase of exchanging. 
Because of the volume contraction, the exchange interaction between 
the f and the conduction electrons is large in CeAgSb$_{2}$. 
This will result in the f electrons to behave like itinerant electrons 
and the narrow f bands to be located at the Fermi level. On the other hand, 
when the exchanging interaction between f and conduction electrons is smaller, 
the occupied 4f orbitals are located near the Fermi level while 
the unoccupied 4f orbitals are at the conduction bands. The quasiparticle 
mass is appropriate to the number of DOS at the Fermi level. So the quasiparticle mass is largely enhanced in CeAgSb$_{2}$. Indeed, it has been shown 
that when the Ce 4f electrons in CeAgSb$_{2}$ are treated as 
localized electrons, 
the 
quasiparticle mass was only enhanced over the band calculation by a factor 
4.8.

\section{Summary}
In this article we showed the results of three different electronic 
band structure calculations.
It shows that the Coulomb potential on Ce 4f orbitals and spin-orbit 
interaction is a key factor 
to understand the electronic and magnetic properties of CeAgSb$_{2}$. 
When the Coulomb 
potential is added to the Ce 4f orbitals, the degeneracy between the different 
f orbits would be lifted and they are split into lower Hubbard bands 
at the Fermi level and unoccupied upper Hubbard bands in the conduction 
band. The exchange interaction between local f electrons and conduction 
electrons play an important role in the heavy fermion characters of them.
And the fully relativistic band structure sheme
shows that spin-orbit coupling splits the 4f states
into two manifolds, the 4f$_{7/2}$ and the 4f$_{5/2}$ multiplet.


\end{document}